\documentstyle[twocolumn,prl,aps]{revtex}
\begin{document}
\twocolumn[\hsize\textwidth\columnwidth\hsize\csname
  @twocolumnfalse\endcsname
\title{Impurity Effects and Spin Polarizations in a Narrow Quantum Hall
System}
\author{Tapash Chakraborty$^\star$}
\address{Max-Planck Institut f\"ur Physik Komplexer Systeme, 
N\"othnitzer Stra{\ss}e 38, D-01187 Dresden, Germany}
\author{K. Niemel\"a and P. Pietil\"ainen}
\address{Theoretical Physics, University of Oulu,
Linnanmaa, FIN-90570 Oulu, Finland}
\date{\today}
\maketitle
\begin{abstract}
The temperature dependence of electron spin polarization for a narrow 
quantum Hall system shows behavior analogous to that of a two-dimensional 
system at major filling factors. At the lowest half-filled quantum Hall 
state for which no two-dimensional analog exists, we find a stable 
spin partially-polarized system. Periodic Gaussian repulsive impurities 
(antidots) in such a system leads to novel spin transitions at 
$\nu=\frac13$ and $\nu=\frac12$ and the pair-correlation functions provide 
clues about nature of different ground states in the system. These results can 
be explored in optical spectroscopy and optically pumped NMR Knight shift 
measurements.

\end{abstract}
  \vskip 2pc ] 

\narrowtext
One of the perhaps most spectacular demonstration of electron correlations in 
nature is the fractional quantum Hall effect (FQHE) \cite{fqhe} for which an 
almost complete picture of the electronic properties at $1/m$ filling of 
the lowest Landau level, $m$ being an odd integer, is available 
\cite{laughlin}. During the rapid developments of our understanding of the 
effect that ensued \cite{book}, one of the fundamental properties of the system 
became well established, i.e., that spin degree of freedom plays a very 
important role in the ground state and elementary excitations in the FQHE 
systems \cite{halperin,spin}. In fact, among the many theoretical predictions 
made within the framework of the incompressible fluid state, only a few have 
received direct experimental support as yet and those include effects based on 
spin polarizations of the two-dimensional electron system (2DES) in the
FQHE regime \cite{spin,charge}. Temperature dependence of the 
spin polarization, calculated recently for the FQHE states and predicted to 
have a non-monotonical behavior for the spin-singlet ground states 
\cite{spinpol}, has also received experimental support \cite{igor}.

When the lowest Landau level is completely filled, the ground state is fully 
spin polarized due to electron-electron interactions \cite{filled}. In 
recent experiments on spin polarizations at and around $\nu=1$, a precipitous 
fall in the spin polarization was observed when either one moves slightly away 
from $\nu=1$ or the temperature is increased at $\nu=1$ \cite{barrett} 
(exceptions also exist, see e.g., Ref.\, \cite{new,igor} where no such 
drop of spin polarization at $\nu\approx1$, or at $\nu\approx\frac13$ was
observed). In this paper we investigate the spin polarizations of electrons
in a narrow quantum Hall wire. We find that most of the features observed 
earlier in two dimensions are preserved in a narrow channel. We also 
demonstrate that the presence of a periodic array of Gaussian scatterers 
(antidot model) has remarkable effects on spin polarizations of the 
incompressible states in a quantum wire. We find abrupt change in spin 
polarizations for a given filling factor as the impurity strength is increased.
In addition, the pair-correlation functions provide a glimpse of the nature
of different ground states in the system.

In our model for the QHE in a narrow channel, we consider a finite 
number of spin polarized electrons interacting via the long-range Coulomb 
potential \cite{narrow} and confined by a potential which is parabolic 
\cite{yoshioka} in one direction and flat in the other. A strong magnetic field
is applied perpendicular to the $xy$ plane. The electrons are confined in a 
cell of length $a$ in $x$ direction and the width of the cell depends on the 
strength of the confining potential $(\frac12m^*\omega_0^2y^2)$ relative to the 
strength of the interactions and also on the length of the cell. We impose a 
periodicity condition in the $x$ direction. For example, we use antiperiodic
boundary conditions for $4n$ electrons so that the non-interacting
ground states have zero total momentum \cite{narrow}.

Electrons are assumed to occupy only the lowest Landau level due to the strong 
magnetic field. The effective magnetic length in the problem is $\lambda=\left(
\hbar/m^*\Omega\right)^{\frac12}$, where $m^*$ is the electron effective 
mass, $\Omega=\sqrt{\omega_0^2+\omega_c^2}$ and $\omega_c=eB/m^*$ is the 
cyclotron frequency. The single-electron wave functions are plane waves in
$x$-direction and oscillator wave functions in $y$-direction centered at
$y_0 = 2 \pi \lambda^2 m/(a \sqrt{1+(\omega_0/\omega_c)^2})$. Here $m$ is the 
momentum quantum number. The corresponding energies, excluding the constant 
Landau level energy are: $E = (2 \pi)^2 (\lambda/a)^2 m$ in units of
$E_0 = (\hbar^2/2m^*\lambda^2)(\omega_0^2/\Omega^2)$.
The Hamiltonian in the lowest Landau level, which
includes contributions from the electron-electron interactions and
the neutralizing background, is numerically diagonalized for a few-electron
system with spin degree of freedom properly included. A phase diagram
is then obtained by plotting the energy gap (energy separation
between the translationally invariant ground state and the lowest excited state)
\cite{narrow} for various values of $a$ and the increasing strength of the
interaction ${\cal E}_c=e^2/\epsilon\lambda$ with respect to the
energy unit $E_0$. We should point out that evaluation of filling 
factors in a 1D system is somewhat tricky. Here the single-particle states 
corresponding to a particular Landau ``level'' are not degenerate. One way
is to calculate the areal electron density and number of fluxes through a unit 
area and determine $\nu$ as the ratio of these two quantities. Alternatively, 
we count the number of occupied states and divide the number of electrons by 
that. Both methods are somewhat arbitrary: one has to choose properly either 
the width of the density profile in the first case (we have used full width at 
half maximum) or, in the second approach, which state should be considered as 
occupied. We have checked that both methods agree reasonably well. The
$\frac13$ FQH state in the present system is also identified from the
momentum distribution function $\langle n(k)\rangle = \langle0|a_k^\dagger a_k|
0\rangle$ by comparing it with that for a Laughlin-like wave function.

\begin{center}
\begin{picture}(200,200)
\put(0,0){\includegraphics{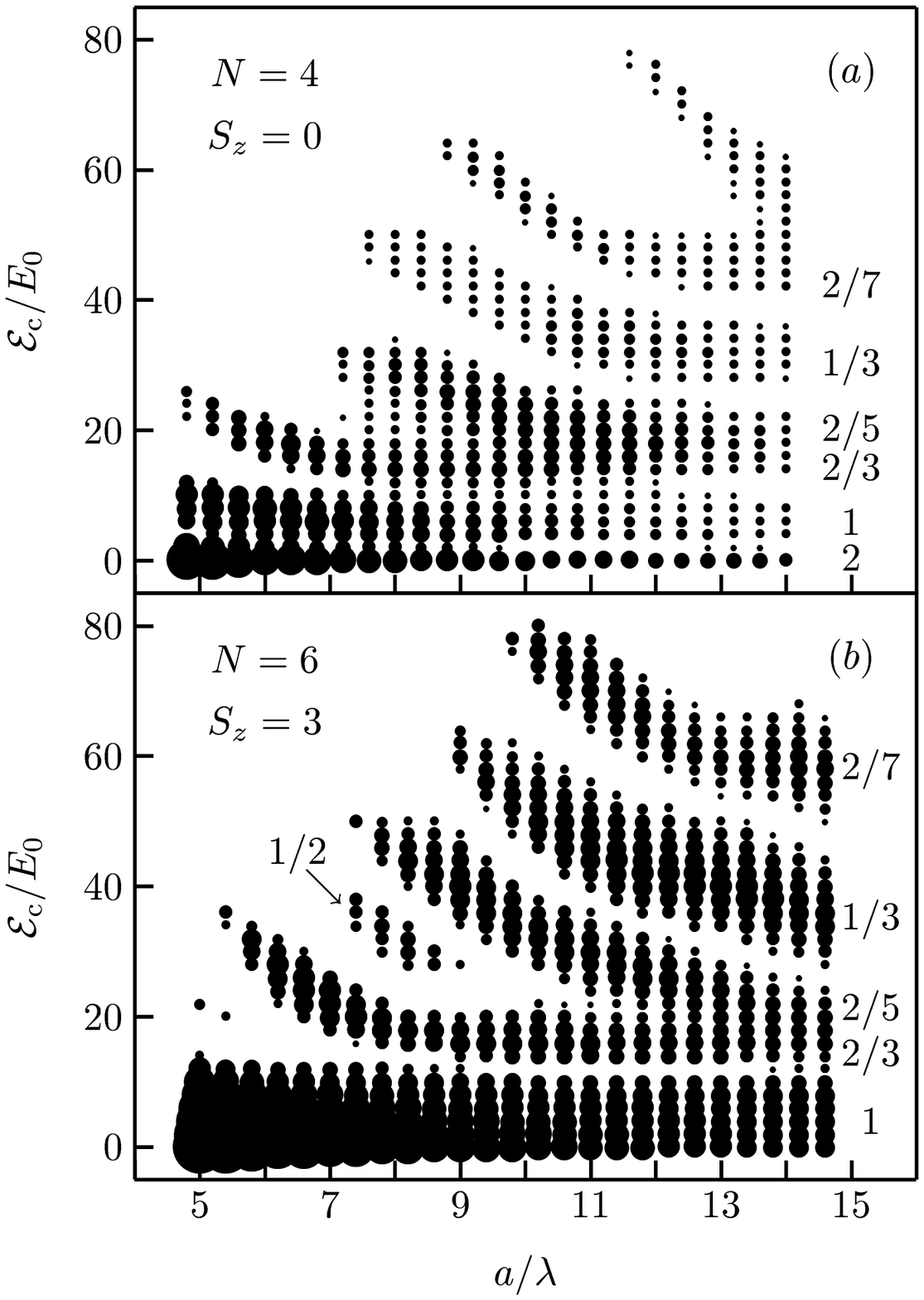}}
\label {fig1}
\end{picture}
\end{center}
\vspace*{5.5cm}

\noindent
FIG. 1: Phase diagram for electrons in a impurity-free narrow channel quantum
Hall system, (a) with and (b) without spin degree of freedom included.
\vspace*{1.0cm}

In Fig. 1, we present the results for the phase diagram, calculated for a
(a) system of six spinless electrons and (b) a system of four electrons with
$S_z=0$ (Zeeman energy not included) and for $\alpha=\omega_0/\omega_c=0.23$ 
which is appropriate for $B=10$ T and $\hbar\omega_0=4$ meV. The area of a 
filled dot is directly proportional to the energy gap. As is evident in the 
figure, several quantum Hall states are stable with large energy gaps in the 
parameter range considered in this work. For the $N=4$ system the $\nu=\frac12$ 
state, though supposed to exist, cannot be resolved in this phase diagram. 
In Fig. 1(a), the $\nu=\frac12$ states are expected to lie between 
$\nu=\frac23$ and $\nu=\frac25$. In general, the energy gaps are larger for 
spinless electrons [Fig. 1(b)] because in the other case there are 
low-lying spin excitations available. 

\begin{center}
 \begin{picture}(200,200)
\put(0,0){\includegraphics{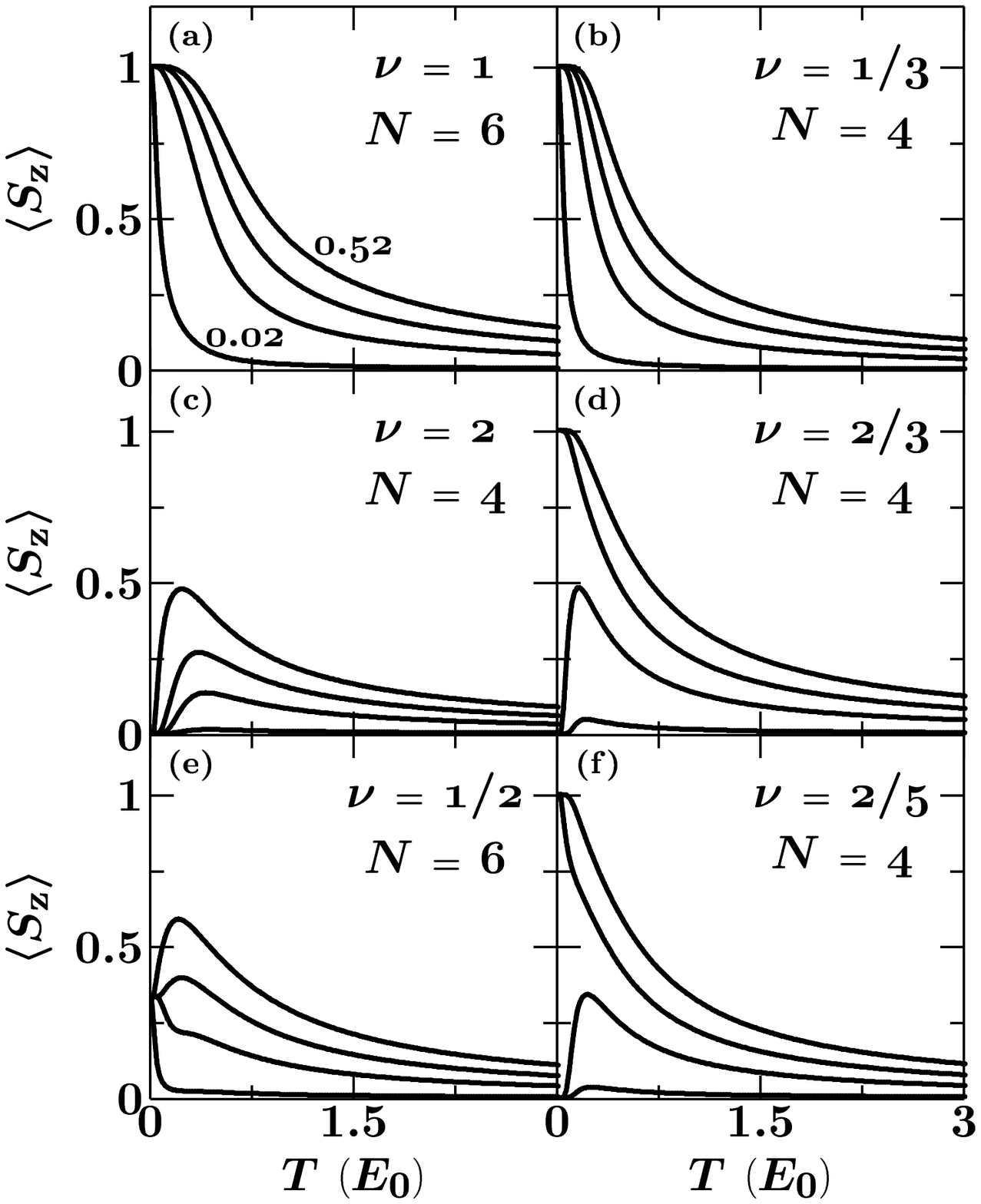}}
\label{fig2}
\end{picture}
\end{center}
\vspace*{4.5cm}

\noindent
FIG. 2: Spin polarization $\langle S_z\rangle$ vs $T$ for $\nu=1,2, \frac23,
\frac12, \frac13$ and $\frac25$ for an impurity-free system.
\vspace*{1.0cm}

The temperature dependence of spin polarization for various filling
factors found in the phase diagram is calculated by a method we developed 
earlier \cite{spinpol}. For the $\nu=\frac12$ results we have employed a
six electron system. The spin polarization $\langle S_z(T)\rangle$ is 
calculated from
$$\langle S_z(T)\rangle\equiv\frac1Z \sum {\rm e}^{-\varepsilon_j/kT}
\langle j|S_z|j\rangle$$
where $Z=\sum_j{\rm e}^{-\beta \varepsilon_j}$ is the canonical partition 
function and the summation is over all states including all possible
polarizations. Here $\varepsilon_j$ is the energy of the state $|j\rangle$
with Zeeman coupling included. A direct measurement of $\langle S_z(T)\rangle$ 
is possible through the NMR Knight-shift measurements and also via optical 
spectroscopic measurements \cite{barrett}. These experiments provide a unique 
probe of spin polarizations in the system.

\vspace*{-0.9cm}
\begin{center}
\begin{picture}(200,200)
\put(0,0){\includegraphics{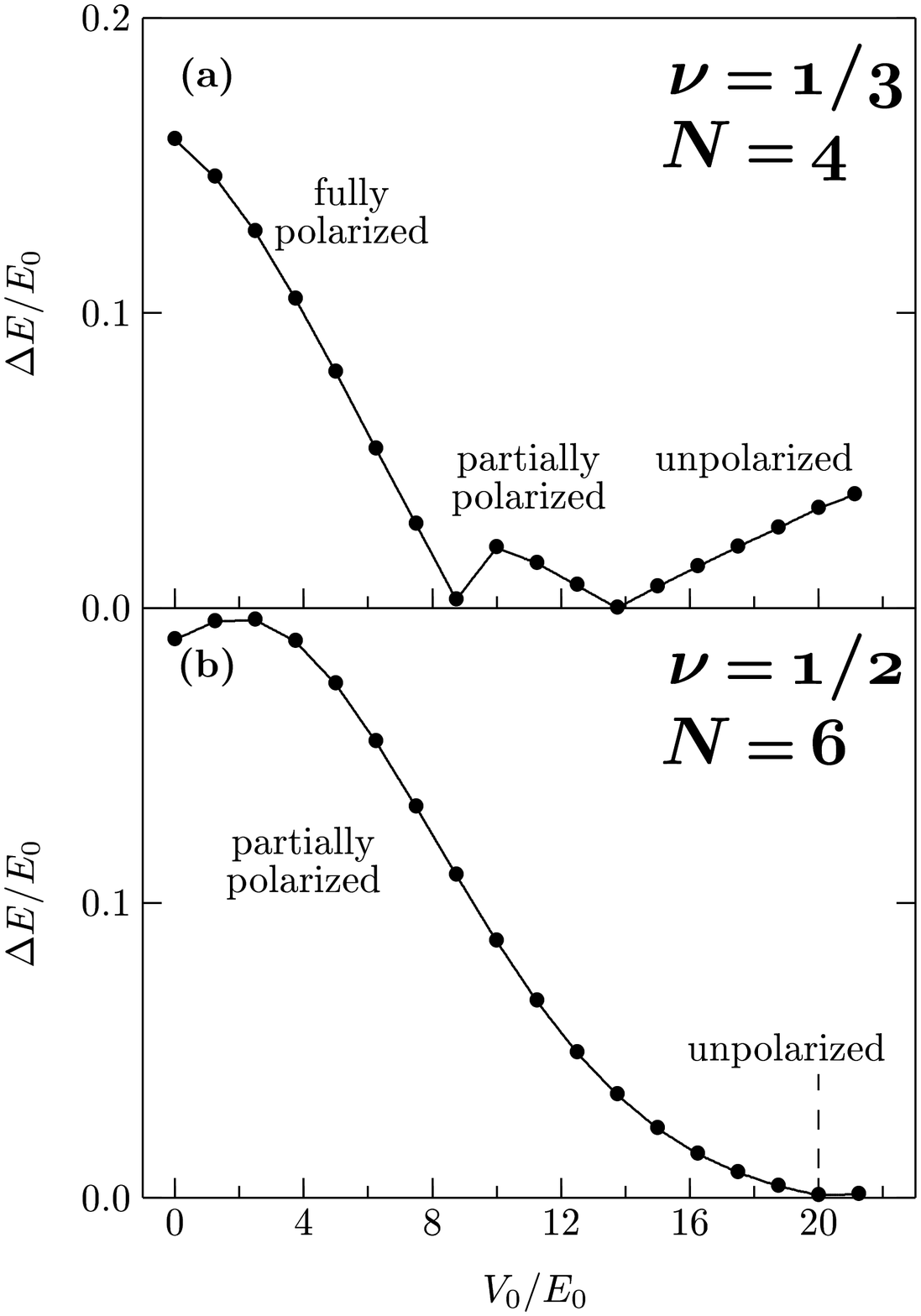}}
\label{fig3}
\end{picture}
\end{center}
\vspace*{6.5cm}

\noindent
FIG. 3: Energy gap at (a) $\nu=\frac13$ and (b) $\nu=\frac12$ in a narrow
channel quantum Hall system as a function of the strength of the Gaussian
repulsive scatterer at the origin and $d=1$.
\vspace*{1.0cm}

Our results for $\langle S_z(T)\rangle$ vs $T$ at $\nu=1,2,\frac23,\frac25,
\frac13$ and $\frac12$ are shown in Fig. 2. In these calculations the
magnetic field was kept fixed at 10T and the $g$-factor is varied 
$(0.02-0.52)$. At $\nu=1$, we find the results to be 
similar to those for the two-dimensional systems \cite{spinpol} and the 
system is fully spin polarized even for very low values of the Zeeman energy. 
Qualitatively similar behavior is also seen at $\nu=\frac13$. In the same way,
$\nu=2$ is a spin-unpolarized state even at the highest value of 
the Zeeman energy considered and $\nu=\frac23$ and $\nu=\frac25$ are
spin-unpolarized states at low Zeeman energies with a non-monotonic temperature 
dependence as predicted for a 2DES \cite{spinpol}. Such a non-monotonical
behavior is observed in experiments on a 2DES \cite{igor}. Clearly, the 
correspondence 
with the spin polarization in a two dimensional system gives us confidence that 
our classification of the QH states in a narrow channel system is essentially 
correct. At $\nu=\frac12$ we find a spin partially-polarized state.

A two-dimensional electron gas with a periodic array of scatterers (antidots) 
is an attractive system to look for the signature of a fermi surface around 
$\nu=\frac12$ \cite{chern,kang} where the well known commensurability 
resonances \cite{weiss} are exploited. In a 2D quantum Hall system, even the 
innocuous $\frac13$ state is known to change its spin polarization in the 
presence of antidot potentials \cite{anti}. We have studied the electronic 
properties of a quantum wire when there is a Gaussian scatterer of the form
$$V^{\rm imp}({\bf r})=V_0\,e^{-\left({\bf r}-{\bf R}\right)^2/d^2}$$
in the cell which, as a result of the boundary conditions, represents a 
periodic array of scatterers. Here $V_0$ is the potential strength, $d$ is the 
width of the potential (in units of magnetic length) and $\bf R$ is the 
impurity position within the cell.

In Fig. 3, we present the energy gap (without including the Zeeman 
contribution) at (a) $\nu=\frac13$ and (b) $\nu=\frac12$ for electrons in a 
narrow channel as a function of the impurity potential strength. 
The energy gap decreases and finally vanishes when the impurity strength is 
increased. With further increase in strength of the impurity potential the gap 
however starts to reappear but with different spin polarizations (and
non-FQH states, as discussed below). At $\nu=1/3$, 
an increase in the strength of the impurity potential seems to cause rapid 
transitions from a spin polarized state to a partially polarized state and 
finally to an unpolarized state. At $\nu=\frac12$, the energy gap also
drops rapidly and the spin state changes from the partially polarized to
the spin unpolarized state, albeit with an extremely small energy gap. 
We should add a cautionary statement here about the $\nu=\frac12$ results
in the presence of a strong impurity potential: the system is too large to 
check the improvement achieved in convergence with respect to increase in basis 
states.

In order to identify the various phases seen at $\nu=\frac13$ in Fig. 3, we 
have calculated the pair-correlation functions for a four-electron system in 
various situations as shown in Fig. 4. Figures 4(a) and 4(b) correspond to
the FQH case in the absence of any antidot potential. Clearly, the extra
structure in $g_{\uparrow\downarrow}(r)$ as compared to 
$g_{\uparrow\uparrow}(r)$ is due to the Pauli principle. Antidot potential has 
only minor effects in the distribution of electrons in the $\frac13$-state as 
long as there is a non-vanishing energy gap in the excitation spectrum 
(Fig. 3). This is evident when one compares Fig. 4 (a),(b) and (c),(d): the 
antidot potential only slightly localizes the electrons. On the other hand, 
there is a dramatic change in the pair-correlation functions after the original 
FQH-gap has vanished i.e., $V_0 = 10$ [Fig. 4 (e) and 4(f)]. The electrons then
are distributed very differently from $\nu=\frac13$ FQH-state and are strongly 
localized along the $y$-axis depending upon the spin polarization of the 
electrons. For $V_0\gg 8$ the states are also non-FQH like.

\vspace*{-4.0cm}
\begin{center}
\begin{picture}(200,200)
\put(0,0){\includegraphics{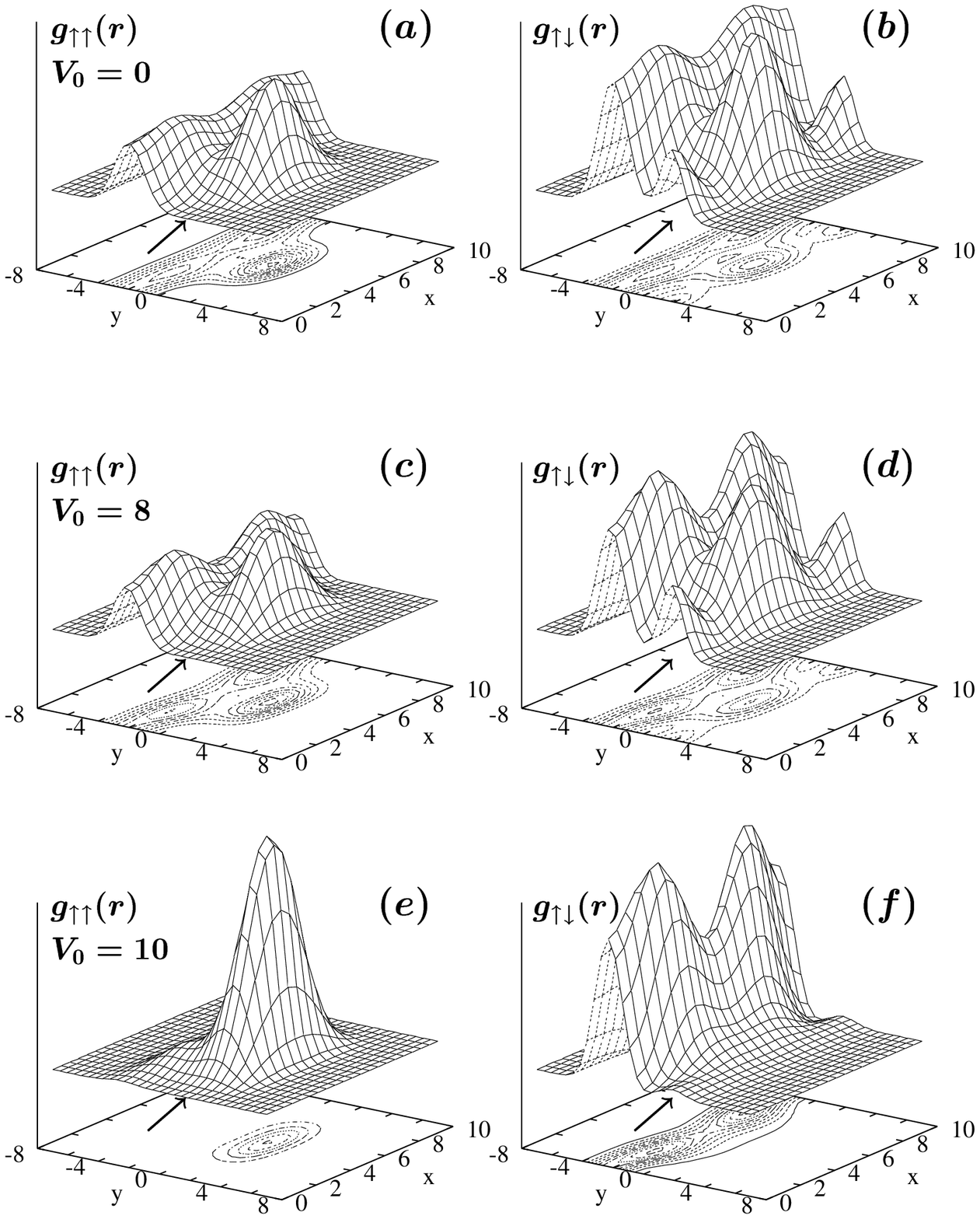}}
\label{fig4}
\end{picture}
\end{center}
\vspace*{10.0cm}

\noindent
FIG. 4: Pair-correlation functions for the $\frac13$-FQH state with $V_0=0$
(a) and (b) and $V_0=8$ (c) and (d). The non-FQH $\frac13$-states
are shown for $V_0=10$ (e) and (f). The arrow indicates the position where
one electron is kept fixed.
\vspace*{1.0cm}
	
To summarize our results, we have studied the temperature dependence of the
spin polarizations of interacting electrons in a narrow quantum Hall system.
We find that there is a clear correspondence with the two-dimensional
behavior at most of the major filling factors. At the half-filled Landau
level we find a spin partially-polarized state. The $\frac13$ FQH state is 
found to be stable against the influence of the impurity potential until the 
energy gap vanishes. The system then goes to a non-FQH state and the impurity
potential strongly localizes the electrons. While in transport measurements 
there are signatures of QH states at $\nu=\frac12$ in a narrow channel 
\cite{timp}, optical spectroscopy and optically pumped NMR Knight shift 
measurements are perhaps more suitable for observation of the temperature 
dependence of spin-polarization at $\nu=\frac12$ and $\nu=\frac13$ in a 
quantum wire as predicted here.

One of us (TC) would like to thank Peter Fulde for his kind hospitality.
KN is supported by the grant of Jenny and Antti Wihuri's foundation.

\end{document}